\documentclass[useAMS,usenatbib]{mn2e}

\usepackage{amssymb,amsmath}
\usepackage{graphicx}
\usepackage{float}
\usepackage{multirow}
\usepackage{aas_macros}
\usepackage{fixltx2e}
\usepackage[usenames]{color}
\usepackage{url}

\newcommand\msun {M$_{\odot}$}
\def\rsun{R$_{\odot}$}
\def\flx{erg cm$^{-2}$ s$^{-1}$}
\def\lum{erg s$^{-1}$}

\def\j1209{J120922+295559}
\def\j0227{J022721+333500}
\def\j0047{J0047}
\def\h{{\it H}-band}
\def\ks{{\it Ks}-band}
\def\vel{km s$^{-1}$}

\title[An RSG counterpart to RX~J004722.4-252051]{Discovery of a red supergiant counterpart to RX~J004722.4-252051, a ULX in NGC 253\thanks{Observations based on ESO programme 093.D-0256}}
\author[M. Heida et al.]
{M. Heida$^{1,2}$, M. A. P. Torres$^{1,2,3}$, P. G. Jonker$^{1,2}$, M. Servillat$^{4,5}$,
\newauthor S. Repetto$^2$, T. P. Roberts$^6$, D. J. Walton$^7$, D.-S. Moon$^8$, F. A. Harrison$^7$\\
$^1$SRON Netherlands Institute for Space Research, Sorbonnelaan 2, 3584 CA Utrecht, the Netherlands\\
$^2$Department of Astrophysics/IMAPP, Radboud University Nijmegen, P.O. Box 9010, 6500 GL Nijmegen, The Netherlands\\
$^3$European Southern Observatory, Alonso de Cordova 3107, Casilla 19001, Vitacura, Santiago 19, Chile\\
$^4$Laboratoire AIM (CEA/Irfu/SAp, CNRS/INSU, Universit\'e Paris Diderot), CEA Saclay, 91191 Gif-sur-Yvette, France\\
$^5$Harvard-Smithsonian Center for Astrophysics, 60 Garden Street, Cambridge, MA 02138, USA\\
$^6$Department of Physics, University of Durham, South Road, Durham DH1 3LE, United Kingdom\\
$^7$Space Radiation Laboratory, California Institute of Technology, Pasadena, CA 91125, USA\\
$^8$Department of Astronomy and Astrophysics, University of Toronto, Toronto, ON M5S 3H4, Canada\\
}

\begin{document}

\maketitle

\begin{abstract}
We present two epochs of near-infrared spectroscopy of the candidate red supergiant counterpart to RX~J004722.4-252051, a ULX in NGC 253. We measure radial velocities of the object and its approximate spectral type by cross-correlating our spectra with those of known red supergiants. Our VLT/X-shooter spectrum is best matched by that of early M-type supergiants, confirming the red supergiant nature of the candidate counterpart. The radial velocity of the spectrum, taken on 2014, August 23, is $417 \pm 4$ \vel. This is consistent with the radial velocity measured in our spectrum taken with Magellan/MMIRS on 2013, June 28, of $410 \pm 70$ \vel, although the large error on the latter implies that a radial velocity shift expected for a black hole of tens of \msun{} can easily be hidden. Using nebular emission lines we find that the radial velocity due to the rotation of NGC 253 is 351 $\pm$ 4 \vel{} at the position of the ULX. Thus the radial velocity of the counterpart confirms that the source is located in NGC 253, but also shows an offset with respect to the local bulk motion of the galaxy of 66 $\pm$ 6 \vel. We argue that the most likely origin for this displacement lies either in a SN kick, requiring a system containing a $\gtrsim 50$ \msun{} black hole, and/or in orbital radial velocity variations in the ULX binary system, requiring a $\gtrsim 100$ \msun{} black hole. We therefore conclude that RX~J004722.4-252051 is a strong candidate for a ULX containing a massive stellar black hole.
\end{abstract}


\section{Introduction}
Ultraluminous X-ray sources (ULXs) are defined as off-nuclear X-ray sources with a luminosity greater than the Eddington luminosity of a 10 \msun{} black hole (BH); in practice a limit for the 0.3 -- 10.0 keV luminosity $\geq 10^{39}$ \lum{} is used (for a review see e.g.~ \citealt{feng11}). The two main scenarios to explain these high luminosities are (1) super-Eddington accretion on to stellar mass BHs or neutron stars (\citealt{begelman02}) and (2) sub-Eddington accretion on to intermediate mass BHs (IMBHs; \citealt{colbert99}).

A growing body of evidence indicates that the group of sources we call ULXs may contain objects of both kinds.
A recent discovery of X-ray pulses from M82 ULX-2 proves that the accretor in the system is a neutron star (\citealt{bachetti14}). Yet this ULX sometimes reaches an X-ray luminosity of $1.8 \times 10^{40}$ \lum, which is clearly orders of magnitude above its Eddington limit. \citet{liu13} use optical spectroscopic observations of the donor star of M101 ULX1 and \citet{motch14} use optical spectroscopic and photometric observations of the donor star of NGC 7793 P13 to calculate a dynamical limit on the BH masses in these systems. In both cases they find the accretor is most likely a stellar mass BH. These recent discoveries support the suggestion of \citet{gladstone09} and others that the peculiar X-ray spectral state, seen in several ULXs with high-quality X-ray data (e.g.~\citealt{bachetti13, sutton13, pintore14}), is an `ultraluminous' state and that many ULXs are stellar mass BHs emitting above their Eddington limit.

However, the brightest ULXs, with luminosities above $10^{41}$ \lum{} (also known as hyperluminous X-ray sources [HLXs]), are still good candidates to host IMBHs (eg. \citealt{farrell09,jonker10,sutton12}). The brightest known ULX and strongest candidate for hosting an IMBH is ESO 243-49 HLX1, with an estimated BH mass of $\sim 20 000$ \msun{} (\citealt{farrell09}). M82 ULX-1 and NGC 2276-3c are two other ULXs that are candidate IMBHs (\citealt{pasham14,mezcua15}). 

The most reliable, model-independent method to measure the mass of the accretors in ULXs is through dynamical mass measurements. The first attempts to do so (\citealt{roberts11,liu12}) used optical emission lines from the accretion discs in these systems. However, these do not show periodic motion. A more promising approach is to measure the radial velocity curve of the donor star, as has been done for several Galactic X-ray binaries (e.g.~\citealt{mcclintock86}).
 
The optical spectra of many ULX counterparts do not show stellar features, as they are dominated by emission from the accretion disc (cf.~\citealt{grise12, tao12a, sutton14}). Some ULXs may however have red supergiant (RSG) donor stars. Since RSGs are intrinsically very bright in the near-infrared (NIR) it is possible to measure their radial velocities spectroscopically out to $\sim 10$ Mpc. An additional advantage is that in the NIR the contribution from the accretion disc is lower than in the optical regime. Also, because of the large orbital separations in such systems, irradiation of the donor star is not expected to play an important role (\citealt{copperwheat07}). Therefore, we performed a photometric survey of nearby ($D < 10$ Mpc) ULXs in the NIR to search for ULXs with RSG donor stars (\citealt{heida14}).

In this paper we present the results of NIR spectroscopic follow-up of the ULX with the brightest candidate NIR counterpart ({\it Ks} = $17.2 \pm 0.5$) in the sample presented in \citet{heida14}: RX~J004722.4-252051 in NGC 253 (hereafter \j0047), at a distance of 3.4 Mpc (\citealt{2009ApJS..183...67D}). The counterpart is also listed in the catalogue by \citet{bailin11} (source ID 47620), with $V = 21.58$ and $I = 19.96$. Given this $V$-band magnitude and the \ks{} magnitude from \citet{heida14}, we derive that the $V - K = 4.4 \pm 0.5$ colour is consistent with that of an RSG, assuming both the $V$ and $K$ magnitudes do not vary strongly in time.

The ULX has been studied in X-rays by for example \citet{barnard10} (NGC 253 ULX1), \citet{sutton13} (NGC 253 XMM2), \citet{pintore14} (NGC 253 X-1) and \citet{wik14} (source 7). It is not a particularly bright ULX, with a peak luminosity in the 0.3 -- 10 keV band of $2.4 \times 10^{39}$ \lum (\citealt{pintore14}), and it has been observed to drop well below the ULX threshold in other observations. Additionally, it shows unusually high variability for a disc-like object (\citealt{sutton13}). The spectrum was classified by \citet{sutton13} as a `broadened disc', an ultraluminous state associated with sources accreting around their Eddington limit (\citealt{gladstone09}). This would imply a mass of $\lesssim 20$ \msun{} for the BH in this system.

\section{Observations and data reduction}
We obtained two spectra of \j0047{} separated by about one year: one with the MMT and Magellan Infrared Spectrograph (MMIRS; \citealt{mcleod12}) on the Magellan Clay telescope at Las Campanas Observatory, and the other with X-shooter (\citealt{vernet11}) on the Very Large Telescope (VLT) UT3 at Cerro Paranal.

\subsubsection*{Magellan/MMIRS} 
The MMIRS spectrum (proposal ID 2013A-CFA-3) was taken on the night of 2013, June 28, with a 2-pixel ($0.4''$) slit, the HK grism and the HK filter. This setup yields a spectrum from 1.25 to 2.45 $\mu$m, with an average spectral resolution ($\frac{\Delta \lambda}{\lambda}$) of R $\approx 1400$. We used an ABBA nodding pattern with a dither offset step of $20''$. We obtained 16 spectra with exposure times of 300 s each, for a total time on source of 4800 s.
The data were reduced through the CfA MMIRS pipeline (v0.9.7RC7-20140301; \citealt{chilingarian13}). This pipeline subtracts the two exposures of one dither pair, performs a flat-field and residual sky correction and rectifies the 2D spectra. The spectra are wavelength-calibrated using the positions of telluric OH-emission lines; the rms of the residuals of the wavelength solution is 0.6 \AA. The dither pairs are then co-added and telluric absorption lines are removed using a telluric standard star observed close in time to and at similar airmass as the target.
These corrected 2D spectra still have both a positive and a negative trace of the object. We use the {\sc starlink} program {\sc figaro} for the further data reduction steps: use the optimal extraction algorithm by \citet{horne86} to extract the spectrum of the positive trace from the pipeline-reduced 2D spectrum, the spectrum of the negative trace from the inverted 2D spectrum, and then add the two to obtain the final spectrum. The nodding pattern unfortunately projected the positive trace of \j0047{} on top of the fainter, extended negative trace of another region of NGC 253. We use a narrow extraction region for the positive trace to minimize the effects of this configuration. We correct the wavelength calibration of the MMIRS pipeline (that is done using wavelengths in vacuum) to match the calibration of our other spectra (that are calibrated using wavelengths in air). 

\subsubsection*{VLT/X-shooter}
The X-shooter spectrum (programme ID 093.D-0256) was taken in service mode on the night of 2014, August 23. X-shooter has three spectroscopic arms that operate simultaneously, providing spectral coverage from the near UV to the near-IR. We used slit widths of $0.8''$ in the UVB arm, $0.7''$ in the VIS arm and $0.6''$ in the NIR arm, resulting in resolutions of 6200, 10600 and 7780, respectively. We used X-shooter in nodding mode, with an ABBA nodding pattern and a nod throw of $5''$. The integration times for the UVB, VIS and NIR arms were 260, 210 and 285 seconds, respectively. The total exposure times for the three arms are 2600, 2100 and 2850 seconds.
 
In the NIR arm, the signal-to-noise ratio (S/N) per spectral resolution element in the {\it J}- and {\it H}-bands is $\sim 10$ in regions that are not affected by atmospheric emission lines. Because the throughput of X-shooter drops rapidly towards the $K$-band, that part of the spectrum is not useful. 

We process the NIR observations of the ULX counterpart and a telluric standard star at similar airmass observed close in time to the target with the X-shooter pipeline workflow in the {\sc Reflex} environment (\citealt{freudling13}). The pipeline produces flat-fielded, sky-subtracted, rectified, wavelength- and flux-calibrated 1D and 2D spectra. The rms amplitude of the residuals of the wavelength solution is 0.1 \AA. The wavelength calibration is done with arcs that are taken during the day, not at the same airmass as the observations. This can introduce a small offset with respect to the real wavelength solution. We check this by comparing the positions of sky emission lines to the line list of \citet{rousselot00} and find an offset of $\approx 0.3$ \AA, that we correct for. We then use the {\sc Spextools}-task {\sc Xtellcor\_general} (\citealt{vacca03}) to correct the spectrum for telluric absorption features.

The \h{} spectra of RSGs contain several strong absorption edges from CO and lines from neutral metals that can be used to derive the effective temperature of the star.The {\it J}-band also contains absorption lines from neutral metals such as Fe, Ti and Mg that can be used to derive abundances, but they are not very sensitive to temperature (\citealt{davies10}). We have verified that these absorption lines are present in the  {\it J}-band, but in the remainder of this paper we focus our analysis on the \h.

In the data from the UVB and VIS arms the S/N of the counterpart is very low ($< 1$). However, there are spatially extended emission lines visible that appear centred on a source at $\sim 5''$ East from the ULX counterpart (not detectable in the NIR arm) that is present in one of the nod positions (the `B' position). From inspection of archival VLT/FORS1 $R$ and $I$-band images taken in the night of 2004, July 23 we know that there is a blue ($V - I = 0$; \citealt{bailin11}) source that lies partially in the slits at that position. The presence of this source makes it impossible to process the data in nodding mode. We follow the X-shooter pipeline manual and use the {\it xsh\_wkf\_stare.oca} file to process the data in stare mode. We extract the emission line spectrum at the position of the counterpart from the exposures without the blue source in the slit (the `A' position), and at a position close to the blue source from the exposures in the `B' nod position. For each exposure, the pipeline produces bias-subtracted, flat-fielded, sky-subtracted, rectified, wavelength- and flux-calibrated 1D and 2D spectra. It also produces 1D and 2D spectra of the subtracted sky emission. For the VIS arm, where there are many strong sky emission lines, we use these sky lines to check the wavelength calibration and find no systematic offset. The rms amplitude of the residuals of the wavelength solution for both the VIS and UVB spectra is 0.06 \AA.

\section{Analysis and results}
\subsection{Cross-correlation analysis}
We are primarily interested in two properties of the candidate counterpart: its spectral type, to give us an indication of the temperature and (via the observed absolute magnitude) the radius of the star, and its radial velocity to confirm that it is a donor star in a binary that belongs to NGC 253 and is not a foreground or background object. Therefore, we cross-correlate our X-shooter spectrum with high S/N spectra of known RSGs of different spectral types (ranging from K1 to M4.5) and metallicities. 

The set of RSG spectra that we use contains one Galactic RSG, CD-60 3621, whose heliocentric radial velocity is $-17.7 \pm 0.4$ km/s (\citealt{mermilliod08}; however, as their individual radial velocity measurements of the star show intrinsic variations from -16.0 to -19.0 km/s, we adopt a larger uncertainty of 3 km/s. All uncertainties quoted in this paper are 1-$\sigma$ errors). Its spectral type is M1.5 I (\citealt{levesque05}). This star was observed with X-shooter as part of the X-shooter Spectral Library on the night of 2010, February 3 (see \citealt{chen14} for a description of the library). We retrieved the pipeline-reduced spectrum from the ESO phase 3 archive. The spectrum has been wavelength calibrated, but not corrected for telluric absorption. The observations were done with the $0.6''$ slit in the NIR-arm, so the resolution of the spectrum is the same as for our X-shooter spectrum (observations for both \j0047{} and CD-60 3621 were done with seeing $> 0.6''$, so the resolution is set by the slit width).

The other RSGs in our comparison sample are those described by \citet{davies13}. They observed 19 RSGs of a range of spectral types in the LMC and SMC with X-shooter. These spectra have been corrected for telluric absorption, but their radial velocities are not as accurately known as that of CD-60 3621. Since these observations were done with a wide ($5''$) slit, they have a lower resolution than our X-shooter spectrum ($R \approx 5000$).

We use the program {\sc molly} for the analysis of the data. After reading the spectra into {\sc molly}, we first use {\it hfix} to calculate and subsequently apply heliocentric corrections. We prepare the spectra for cross-correlation following the recommendation in the {\sc molly} user guide\footnote{\url{http://deneb.astro.warwick.ac.uk/phsaap/software/molly/html/USER_GUIDE.html}}, first normalizing with a second-order polynomial fit and then subtracting a fourth-order spline fit to the normalized spectra. We resample the spectra to a common velocity spacing (10.85 km s$^{-1}$ pix$^{-1}$ for our X-shooter spectrum, 118.4  km s$^{-1}$ pix$^{-1}$ for our MMIRS spectrum) with {\it vbin} and then run {\it xcor} to calculate the cross-correlation functions of \j0047{} with the comparison sample. {\it xcor} also calculates the radial velocity offset from the cross-correlation function by fitting a parabola to the three peak pixels. This is a generally robust estimate but it underestimates the uncertainty on the radial velocity. To get a better estimate of the uncertainty we fit a polynomial plus Gaussian curve to the cross-correlation functions using {\it mgfit}. As a sanity check we also calculate the cross-correlation functions of CD-60 3621 with the LMC/SMC RSGs. The parts of the spectrum of \j0047{} that are most affected by noise from strong telluric emission lines are masked out during the cross-correlation.

\begin{figure*}
\hbox{
\includegraphics[width=0.5\textwidth]{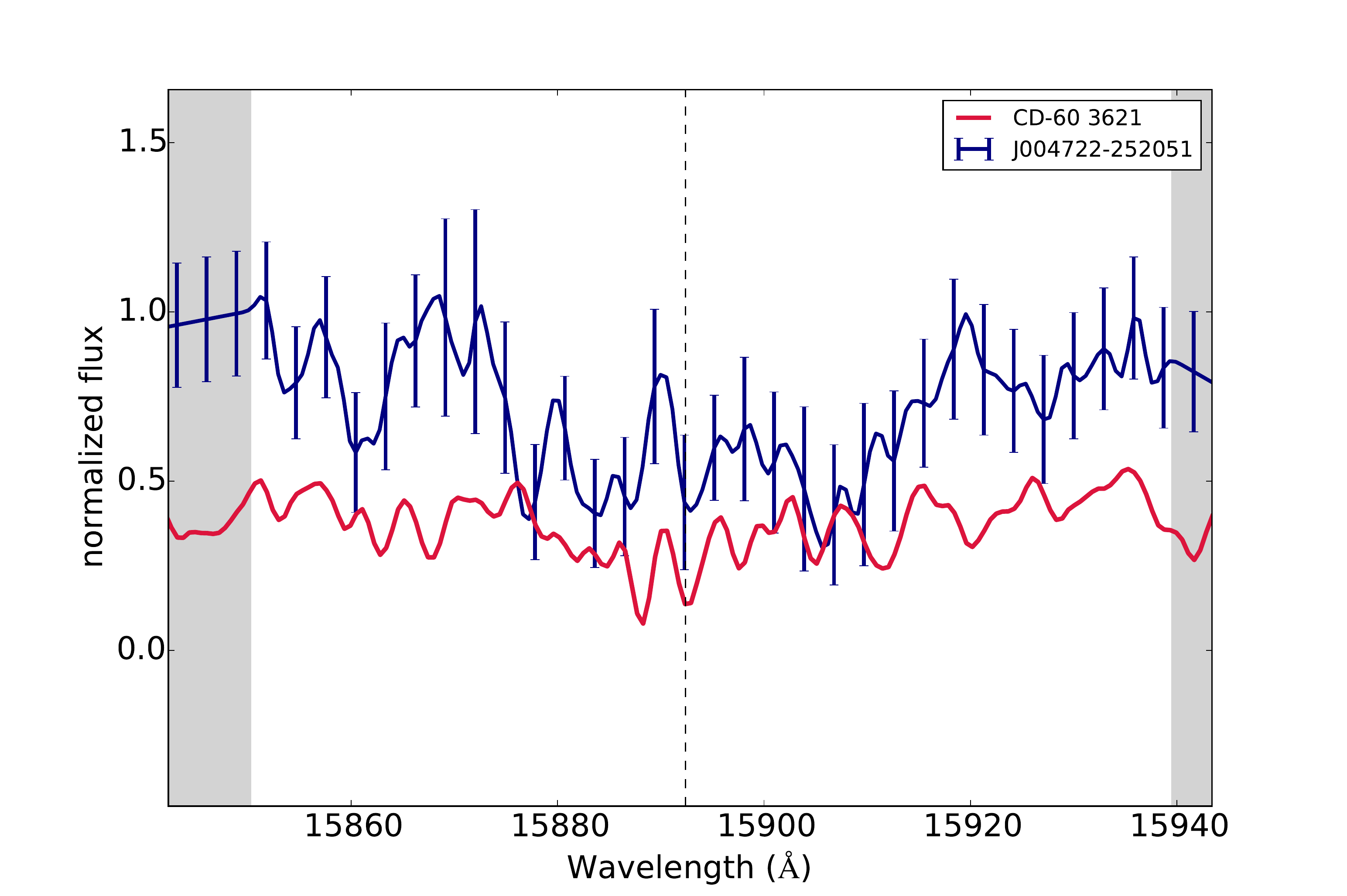}
\includegraphics[width=0.5\textwidth]{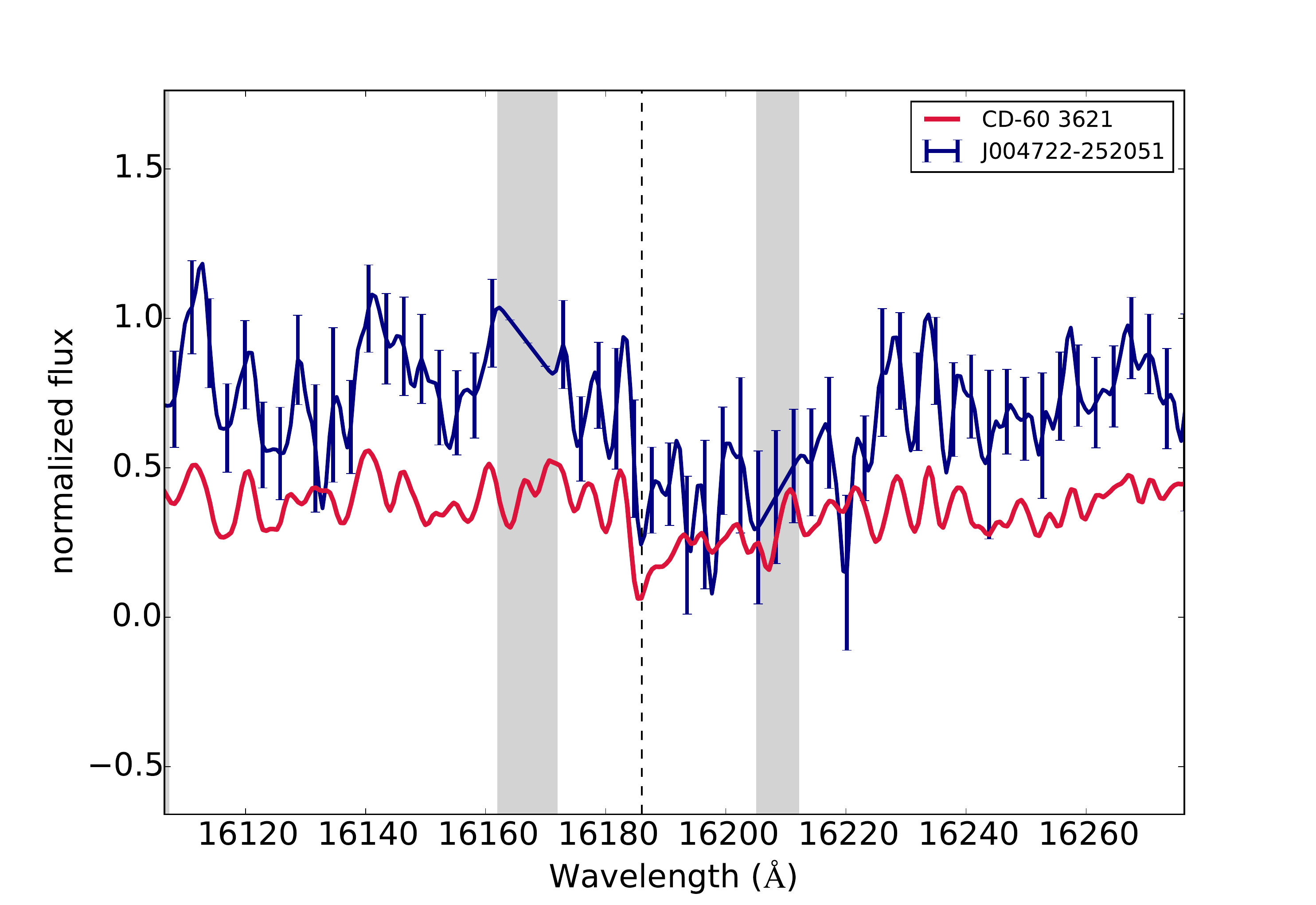}
}
\hbox{
\includegraphics[width=0.5\textwidth]{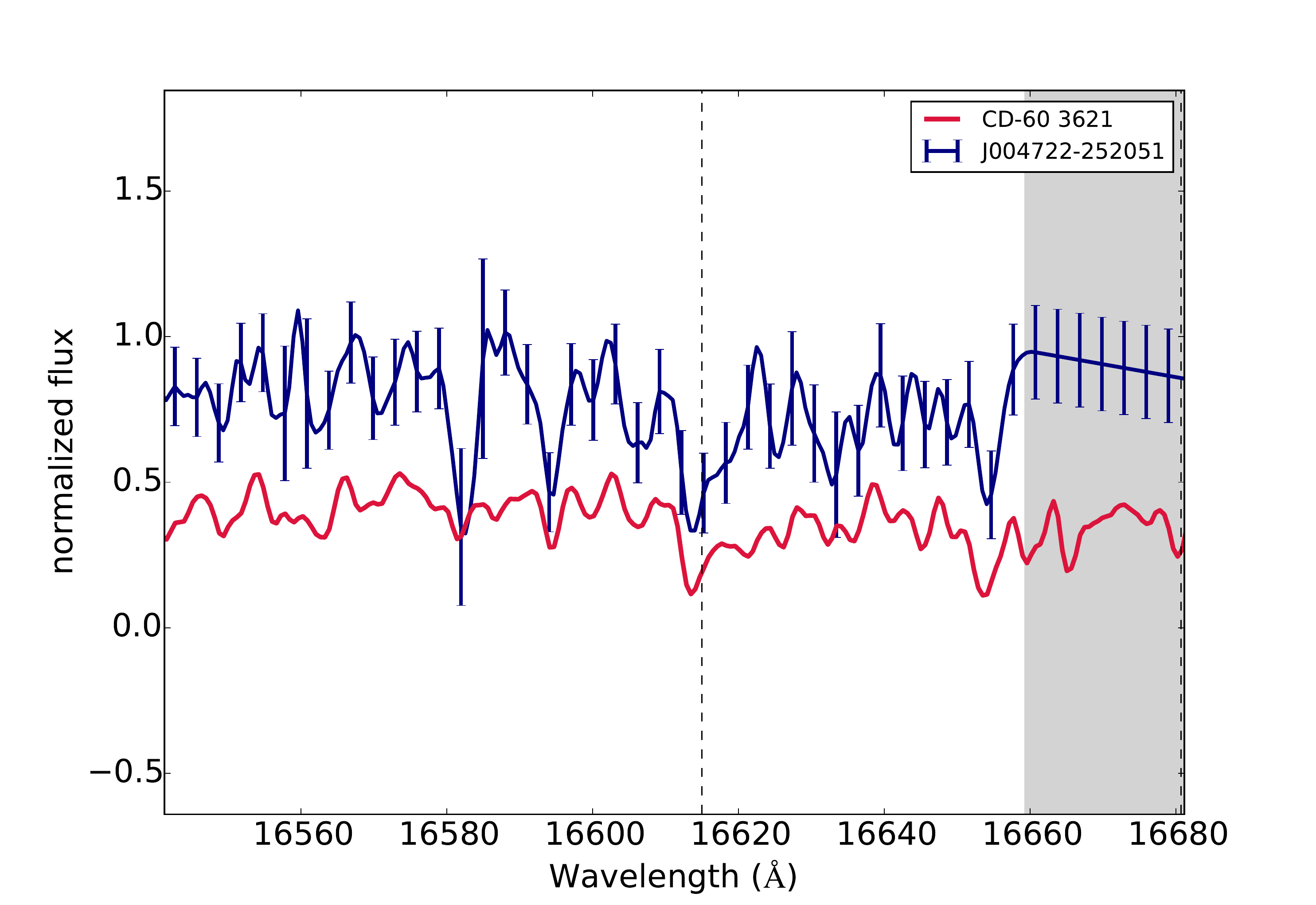}
\includegraphics[width=0.5\textwidth]{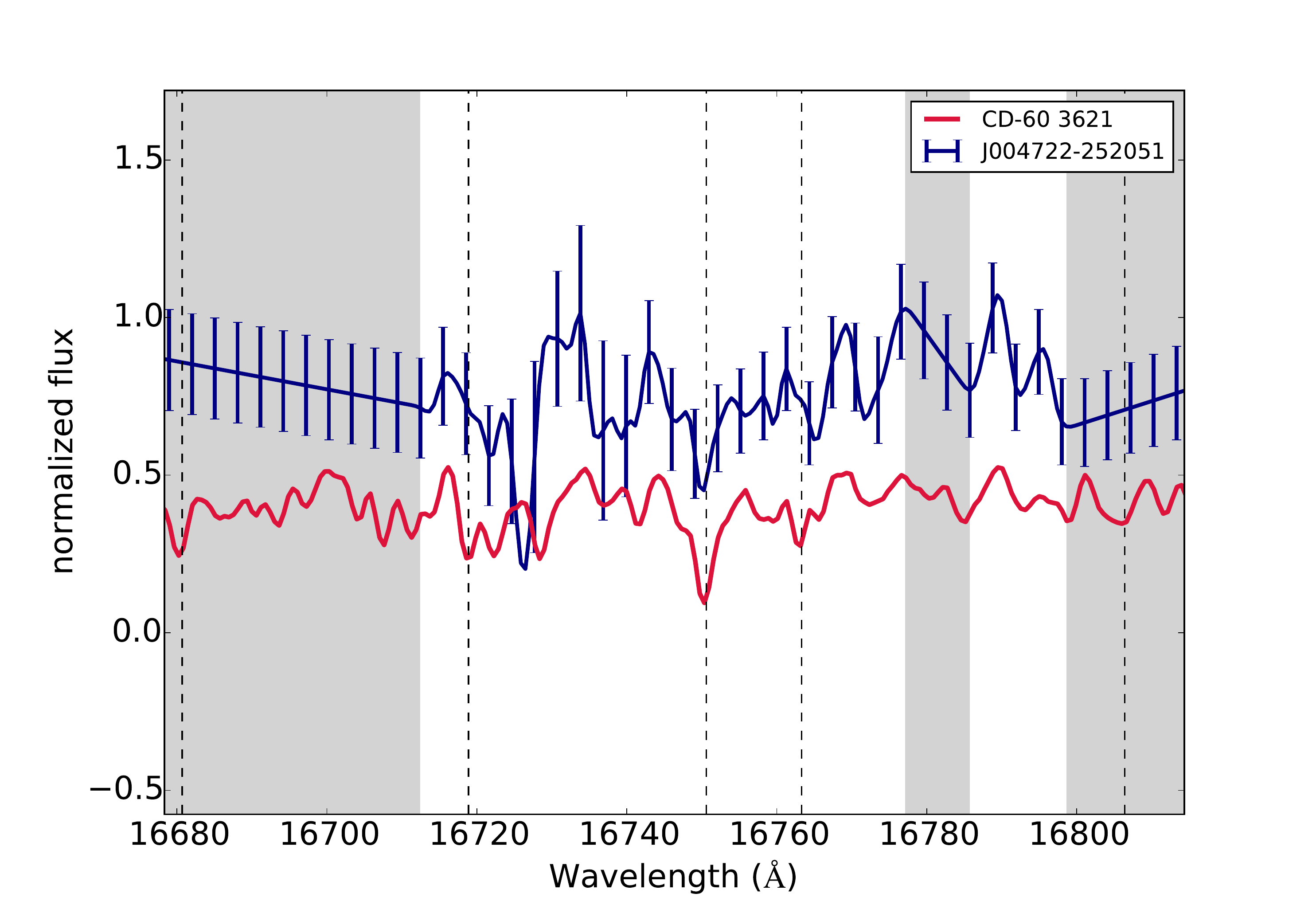}
}
\hbox{
\includegraphics[width=0.5\textwidth]{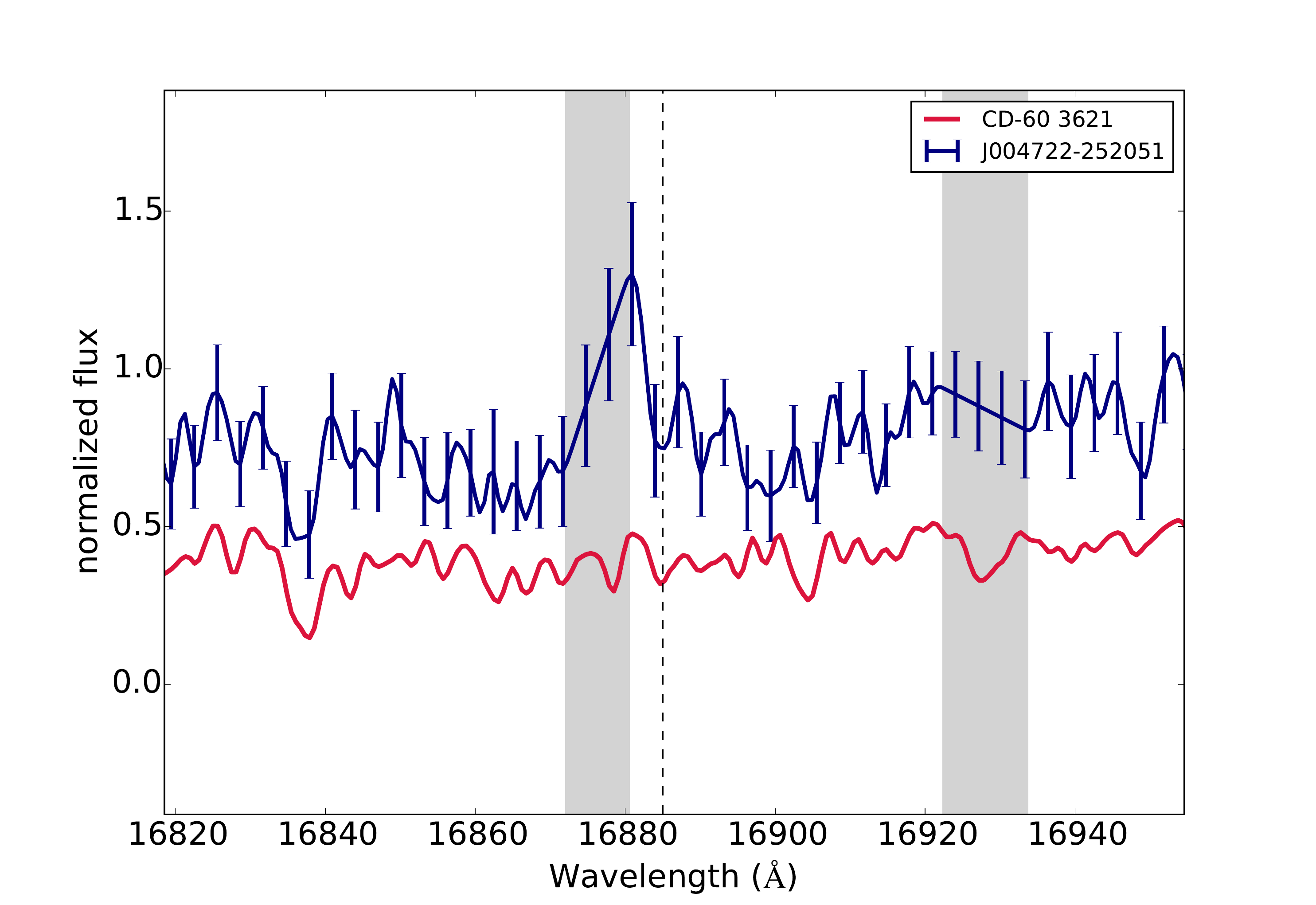}
\includegraphics[width=0.5\textwidth]{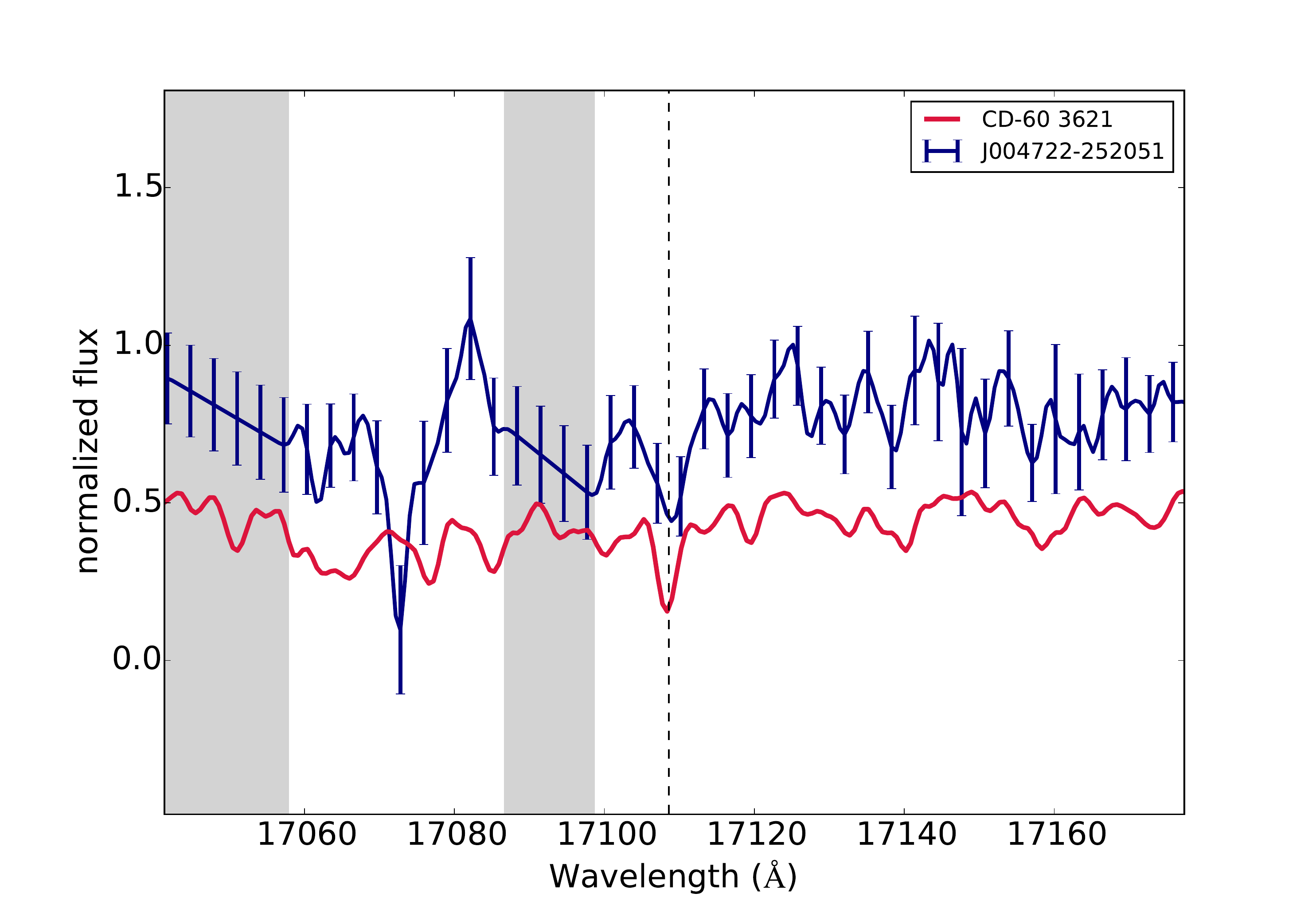}
}
\caption{Sections of the \h{} X-shooter spectrum of \j0047{} (upper, blue line) and the X-shooter spectrum of CD-60 3621 (lower, red line). Both spectra are normalized and smoothed to the resolution of X-shooter. For presentation purposes only, we interpolated the spectrum of \j0047{} over areas that are strongly affected by telluric emission (grey shaded areas). Error bars are plotted for every fifth data point. The spectrum of CD-60 3621 is shifted vertically by $-0.5$. The \j0047{} spectrum is shifted by $-417$ \vel{} and the CD-60 3621 spectrum is shifted by 17 \vel. The dashed vertical lines indicate the positions of atomic and molecular lines (left to right, top to bottom: Si {\sc i} (1.59 $\mu$m), CO bandhead (1.62 $\mu$m), CO bandhead (1.66 $\mu$m), Al {\sc i} triplet (1.672, 1.675, 1.676 $\mu$m), OH (1.69 $\mu$m), Mg {\sc i} (1.71 $\mu$m). 
}\label{fig1}
\end{figure*}

The cross-correlation between the \h{} X-shooter spectrum of \j0047{} and CD-60 3621 yields a velocity shift of $434.4 \pm 1.0$ km s$^{-1}$. This uncertainty is the statistical error on the determination of the position of the peak of the cross-correlation function. The total uncertainty includes the uncertainty on the wavelength calibration of X-shooter (2 km s$^{-1}$ for both spectra) and the uncertainty in the radial velocity of CD-60 3621 (3 km s$^{-1}$). Quadratically adding all these uncertainties and adding the radial velocity of CD-60 3621 gives a velocity shift of $417 \pm 4$ km s$^{-1}$. 
The cross-correlations of \j0047{} with the SMC/LMC sample yield consistent results. 

We obtain the highest values of the cross-correlation function for CD-60 3621 and LMC-136042 (the latter is an M3-type RSG). In addition, following the procedure described by \citet{origlia93}, we measure the equivalent widths of the CO (1.62 $\mu$m) bandhead and the Si I (1.59 $\mu$m) absorption line using the {\it light} command in {\sc molly}  (see Figure \ref{fig1} where parts of the spectra are plotted, shifted to zero radial velocity with respect to the heliocentric frame). The ratio of these lines is sensitive to the temperature of the star. We find a ratio of $\log(W_{1.62}/W_{1.59}) = 0.15 \pm 0.1$, which is consistent with an effective temperature in the range 3000 -- 4000 K [\citealt{forster00}; with the caveat that our spectrum has a higher resolution than those of \citet{forster00} and \citet{origlia93}, and it is not known if this will affect the line ratio].

The cross-correlation functions for the SMC RSGs all have lower values than those for the LMC RSGs. From this we conclude that \j0047{} is an early M-type RSG, with a metallicity closer to that of the LMC and Milky Way RSGs than to that of those in the SMC, which have a lower metallicity. This is not unexpected, as abundance measurements of H {\sc ii} regions in NGC 253 yield values for 12 + log(O/H) in the range 8.5 -- 9.0 (\citealt{webster83}), comparable to abundance ratios found in H {\sc ii} regions in the Milky Way (e.g.~\citealt{balser11}). Values found for H {\sc ii} regions in the LMC and SMC are $8.4 \pm 0.12$ and $7.98 \pm 0.09$, respectively (\citealt{pagel78}).

The Magellan/MMIRS spectrum of \j0047{}, due to its much lower resolution and lower S/N, can not be used to constrain the spectral type of the RSG. Therefore, we cross-correlate this spectrum only with CD-60 3621, resampled to the velocity resolution of the MMIRS spectrum. We initially mask only those pixels that are strongly affected by residuals from telluric emission lines. This leaves 169 pixels (from a total of 463) for the cross-correlation. The peak of the cross-correlation function is found at a shift of $3.5 \pm 0.2$ pixels, or $410 \pm 20$ km s$^{-1}$. As a consistency check we also cross-correlate the spectra using only regions where we could visually identify absorption lines. This leaves us with 63 pixels. The shift found in this way is consistent with the first: $3.6 \pm 0.3$ pixels, or $430 \pm 30$ km s$^{-1}$. Correcting these shifts for the radial velocity of CD-60 3621 gives values of $400 \pm 20$ and $410 \pm 30$ km s$^{-1}$, respectively. However, the uncertainty on the position of the peak of these cross-correlation functions is calculated assuming they have a Gaussian profile. This is the case for the X-shooter data but not for the MMIRS data. Therefore, the error on the radial velocity of the MMIRS spectrum is underestimated.

\begin{figure*}
\includegraphics[width=\textwidth]{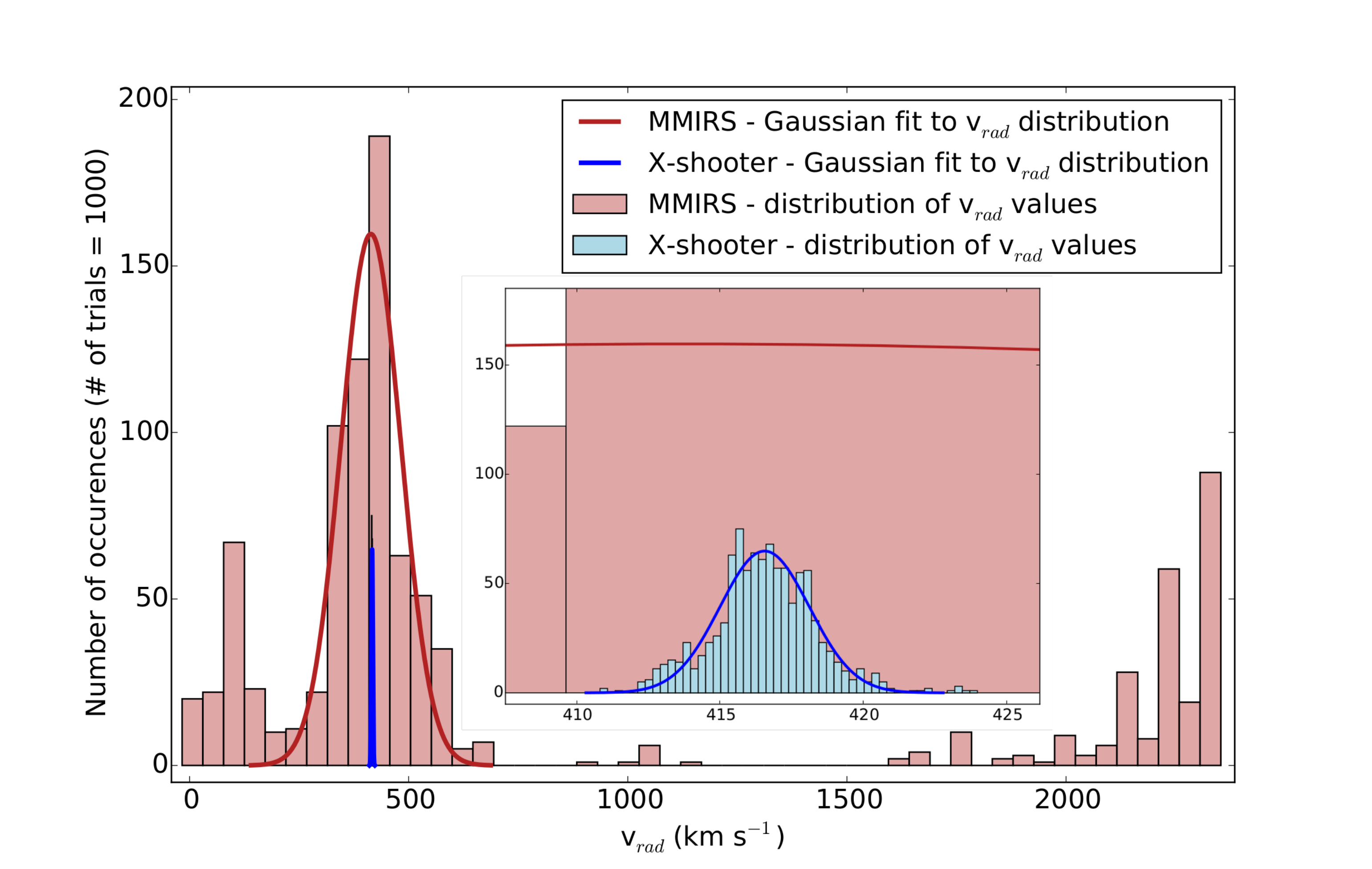}
\caption{The histograms show the distribution of radial velocities obtained by cross-correlating 1000 bootstrapped copies of the MMIRS spectrum (red) and the X-shooter spectrum (blue) with CD-60 3621. The systemic velocity of CD-60 3621 of 17 \vel{} is already subtracted. The best-fitting Gaussian curves to the data are also plotted (red/blue lines). Inset is a zoom-in of the image around 415 \vel{} to show the X-shooter data more clearly.} \label{fig:hist}
\end{figure*}

To gain more insight in the uncertainties on the radial velocities, we perform a bootstrap analysis. We use the {\sc molly} command {\it boot} to produce 1000 resampled copies of the MMIRS and X-shooter spectra. We then cross-correlate these with the spectrum of CD-60 3621 using {\it xcor}, and fit a Gaussian to the resulting distribution of radial velocities. The width of this Gaussian is a good measure of the uncertainty on the radial velocity. After correcting for the radial velocity of CD-60 3621 we find a value of $416.6 \pm 1.6$ \vel{} for the X-shooter spectrum. As we expected, this value is consistent with the one we obtained directly from the cross-correlation. For the MMIRS spectrum, we find a radial velocity of $410 \pm 70$ km s$^{-1}$ if we use the 169-pixel mask (see Figure \ref{fig:hist}), and $410 \pm 30$ km s$^{-1}$ with the 63-pixel mask. We conservatively adopt that the radial velocity of \j0047{} as measured from the MMIRS spectrum is $410 \pm 70$ km s$^{-1}$.

\subsection{Nebular emission lines}
\begin{figure}
\includegraphics[width=0.5\textwidth]{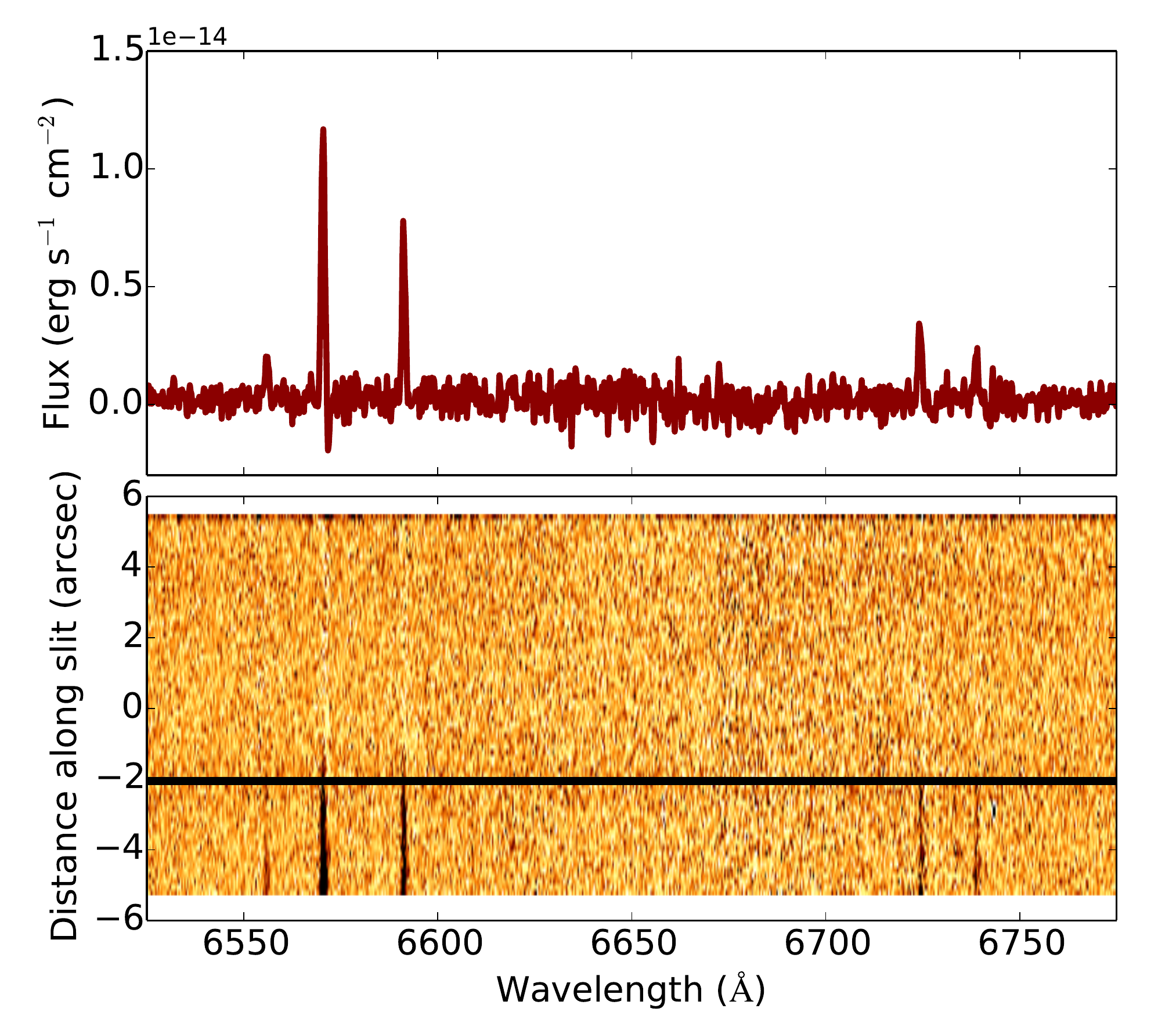}
\caption{Part of the VIS spectrum with the [N {\sc ii}]$\lambda 6548$, H$\alpha$, [N {\sc ii}] $\lambda 6583$ complex and the [S {\sc ii}] $\lambda\lambda 6716, 6731$ doublet. Upper figure: the flux-calibrated spectrum extracted at the position of the ULX. Lower figure: the 2D image of the spectrum. Here it can clearly be seen that the emission lines are extended. The black horizontal line indicates the position of the RSG (not visible) on the detector.}\label{visspec}
\end{figure}

Although the UVB and VIS spectra have very weak continua (UVB: S/N $< 1$, VIS: S/N $\approx 1$) there are several emission lines visible in the data. In the UVB data, these are the [O {\sc ii}]$\lambda\lambda 3727,3729$, H$\beta$ and [O {\sc iii}]$\lambda\lambda 4956, 5007$ lines. In the VIS spectrum we detect [N {\sc ii}]$\lambda 6548$, H$\alpha$, [N {\sc ii}] $\lambda 6583$ and the [S {\sc ii}] $\lambda\lambda 6716, 6731$ doublet. Inspection of the 2D spectra reveals that these lines do not fill the whole slit but are extended in the direction of --- and peak around --- the bright source that dominates the unused nod position B (this source is located $\sim 5''$ East of the RSG, but is not visible in our NIR image due to its blue colour; for the 1 and 2D VIS spectra see Figure \ref{visspec}).
After reading the spectra of the single exposures into {\sc molly} and applying {\it hfix}, we average them using the task {\it average}. We then measure the radial velocity of the emission line region in the UVB and VIS spectra separately by using the {\it mgfit} function to fit a set of Gaussian curves with a common velocity offset to the above-mentioned lines. The average radial velocity we find in this way is $351 \pm 4$ \vel.

\begin{figure*}
\includegraphics[width=\textwidth]{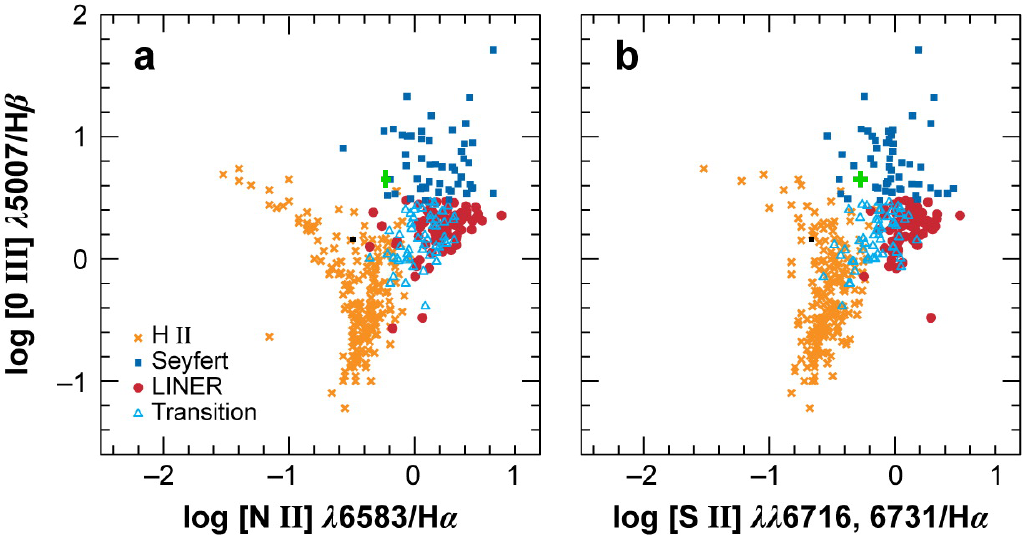}
\caption{The position and error bars of the line ratios in diagnostic diagrams, indicated by black squares (for the spectrum extracted close to the blue source) and green pluses (for the spectrum extracted at the position of the ULX; figure adapted from \citealt{ho08}). Close to the blue source the line ratios are fully consistent with a H {\sc ii} region, while at the position of the ULX the line ratios are more like those seen in Seyferts, indicating that the ULX might play a role in ionizing the part of the nebula that surrounds it.}\label{dds}
\end{figure*}

From the fits to the spectra we also obtain line ratios of log([O {\sc iii}]/H$\beta$), log([N {\sc ii}]/H$\alpha$) and log([S {\sc ii}]/H$\alpha$). These ratios are plotted in Figure \ref{dds}, adapted from \citet{ho08}. Close to the blue source the line ratios are fully consistent with a H {\sc ii} region, while at the position of the ULX the line ratios are more like those seen in Seyferts.

The He {\sc ii} $\lambda 4686$ line, which acts as a photon counter for photons in the 54 -- 200 eV band (\citealt{pakull86}), has been observed in several ULX nebulae (cf.~\citealt{pakull02,kaaret09,moon11,gutierrez14}). We do not detect it in our X-shooter spectrum, with a $2-\sigma$ (99.7 \% confidence) upper limit of $8.8 \times 10^{-17}$ \flx{} (corresponding to a luminosity of $1.2 \times 10^{35}$ \lum).

\subsection{Probability of chance superposition}
\begin{figure*}
\includegraphics[width=\textwidth]{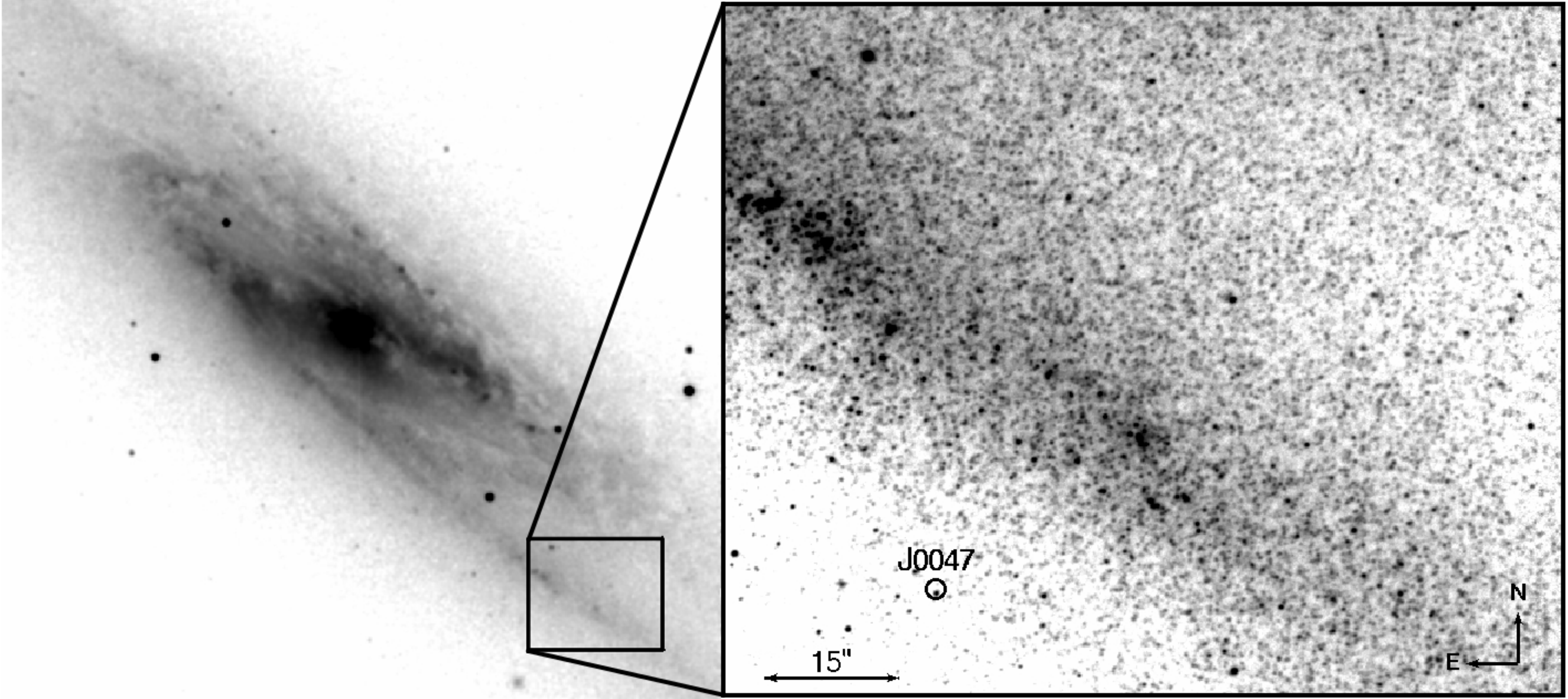}
\caption{NIR image of the region around \j0047. Left: 2MASS image of NGC 253. Right: Our VLT/ISAAC \ks{} image with the X-ray position of the ULX indicated by the black circle (radius $1.1''$, 95\% confidence).}\label{finder}
\end{figure*}

\j0047{} is located close to one of the spiral arms of NGC 253 (see Figure \ref{finder}). In such an environment, the probability of a chance superposition of an unrelated object may not be negligible.
Using the  {\sc iraf} task {\it daofind} we search for point sources in our VLT/ISAAC \ks{} image (\citealt{heida14}) that are equally bright or brighter than the apparent magnitude of the candidate RSG ({\it Ks} = 17.2 $\pm$ 0.5). We find that there are 50 such sources in our image, that has a total area of 7200 arcsec$^2$. The error circle around the X-ray position of the ULX has a radius of $1.1''$, or an area of 3.8 arcsec$^2$ (see \citealt{heida14} for the calculation of the size of the error circle and details of the ISAAC NIR image). This means the probability of finding a source as bright as the RSG in the error circle by chance superposition is 2.6\%. However, this image includes part of a spiral arm of NGC 253 while the ULX is located just outside this arm, where the point source density is lower. Therefore, 2.6\% is a conservative upper limit. Due to the inhomogeneous distribution of stars, the value for the chance superposition probability depends strongly on the part of the image that is selected for the calculation. Therefore, it is not possible to find an accurate value for the chance superposition probability.
With an upper limit of 2.6\% we cannot exclude the possibility that the RSG and ULX are unrelated. Only a robust measurement of radial velocity variations will prove whether the RSG is orbiting the compact object responsible for the ULX emission.

\section{Discussion and conclusions}
We obtained low- and medium-resolution \h{} spectra of the counterpart of \j0047, a ULX in NGC 253. From a cross-correlation with spectra of RSGs in the Milky Way, LMC and SMC we conclude that the candidate counterpart is most likely an early M-type supergiant. 

In the near-UV and visible parts of our X-shooter spectrum we detect several emission lines with line ratios that are in between those of H {\sc ii} regions and Seyferts (see Figure \ref{dds}). The 2D profile of the emission lines shows that they are extended in the direction, and peak at the position, of a bright and blue optical source at $\sim 5''$ (corresponding to $\sim 85$ pc at the distance of NGC 253) from the RSG. This source is not visible in our NIR image but was reported by \citet{bailin11} to have M$_V = -8.1$ and $V - I = 0$, consistent with a late B or early A supergiant. Close to this blue source the ratios of the emission lines are different from the ratios found at the position of the ULX, and fully consistent with a H {\sc ii} region.
We do not detect the He {\sc ii} $\lambda 4686$ line that is indicative of X-ray illuminated nebulae, with a $2-\sigma$ upper limit of $8.8 \times 10^{-17}$ \flx{} (corresponding to a luminosity of $1.2 \times 10^{35}$ \lum). For comparison, \citet{moon11} report a luminosity in this line of $\sim 8.8 \times 10^{34}$ \lum for M81 X-6, a ULX with an X-ray luminosity similar to \j0047. 
We propose that the nebula is excited both by young stars and the ULX, tracing a region with recent star formation where also the RSG was formed. A detection of - or a stronger limit on - the He {\sc ii} $\lambda 4686$ line could (dis)prove this scenario.

We use these lines to measure the projected radial velocity of NGC 253 at the position of the ULX and find a value of $351 \pm 4$ \vel. This is compatible with the value of $350 \pm 30$ \vel{} reported by \citet{hlavacek11}, who measured the rotation curve of the galaxy using H$\alpha$ observations.
The radial velocity of the counterpart, as measured from our VLT/X-shooter spectrum taken in August 2014, is $417 \pm 4$ \vel. This proves that it is not a foreground or background object but is located in NGC 253, confirming its absolute \ks{} magnitude of $-10.5 \pm 0.5$ and its identification as an RSG (\citealt{heida14}). This absolute magnitude is too bright for asymptotic giant branch (AGB) stars, that have absolute $K$-band magnitudes between $-6.4$ and $-8.2$ (\citealt{knapp03}).
This makes \j0047{} a very strong candidate for a ULX with an RSG donor star. The ultimate proof for this would be the detection of periodic radial velocity variations in the RSG. 

\begin{figure}
\includegraphics[width=0.5\textwidth]{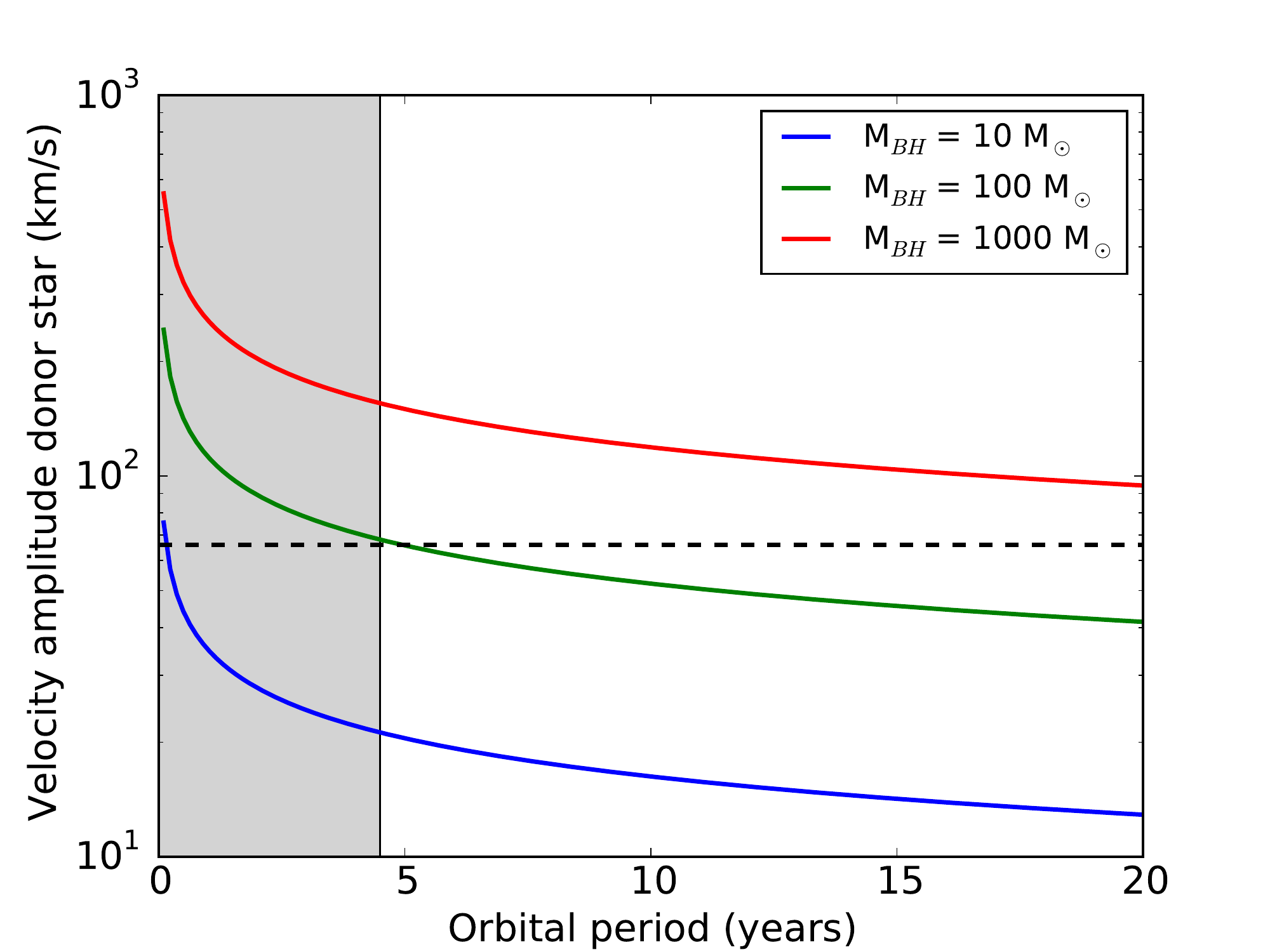}
\caption{The radial velocity amplitude for a 10 M$_\odot$ mass donor in orbit around a 10, 100, or 1000 M$_\odot$ black hole (lower, middle and upper line, respectively) as a function of orbital period. A system inclination of 60$^\circ$ is assumed. Assuming a Roche lobe filling RSG, the expected orbital period is between 4.5 and 20 years. Smaller donor radii give shorter orbital periods; a higher donor mass requires a higher BH mass to reach the observed velocity. The black dashed line indicates a velocity of 66 \vel; the hashed area is ruled out based on the necessary radius of the RSG.}\label{veloamp}
\end{figure}

With an effective temperature in the range of 3000 -- 3900 K (spectral types M0--3, \citealt{tokunaga00}), the expected radius of the RSG based on its absolute \ks{} magnitude of $-10.5$ is $\approx 600-1600$ \rsun. Assuming Roche lobe filling RSG with a mass of 10 \msun{}, the orbital period expected for such an RSG orbiting a BH is 4.5 -- 20 years.
The apparent radial velocity amplitude depends on this period as well as on the mass ratio and the inclination of the system. For instance, for an inclination of 60$^{\circ}$, this velocity amplitude varies from $\sim 20$ \vel{} in the case of a stellar mass BH, to hundreds \vel{} for an IMBH (see Figure \ref{veloamp}). 
We do not detect a significant difference in radial velocity between our June 2013 and August 2014 spectra ($\Delta$v = 10 $\pm$ 70 \vel). The large uncertainty is due entirely to the June 2013 low resolution, low S/N MMIRS spectrum, and prevents us from putting strong limits on the mass of a possible BH companion. We calculate the probability of detecting a velocity shift larger than 80 \vel{} (our 1-$\sigma$ upper limit on the velocity shift) for a range of black hole masses and orbital periods, taking into account that the orbital phase is unknown. For a system inclination of 60$^{\circ}$, we find that if the orbital period is 4.5 years, we have a $< 50\%$ chance of detecting a velocity shift if M$_{\textrm{BH}} < 100$ \msun. If the orbital period is more than 10 years we would not detect RV variations even if the RSG orbits a 1000 \msun{} BH. 

However, we do measure a radial velocity offset of $66 \pm 6$ \vel{} between the RSG and its surroundings (as traced by the emission lines).  We consider three possible explanations for this offset:

{\bf 1}: The RSG and the X-ray source are unrelated, and the RSG is a single runaway star. 
In this scenario the ULX has another, fainter donor star that we do not detect. \citet{eldridge11} calculated the runaway fraction of RSGs in a simulated stellar population; at solar metallicity, they find that 7.2\% of RSGs have a space velocity $> 30$ \vel{} (their table 4). Of those runaway RSGs, $\sim 10\%$ have a space velocity $> 60$ \vel{} (the 1-$\sigma$ lower limit to our velocity offset; their figure 2). We combine this with the upper limit of 2.6\% that we find for the chance superposition probability of the RSG with the ULX. Using these conservative estimates (since we only measure the radial velocity of the RSG and the space velocity is likely larger) we find that the probability of a chance superposition of the ULX with a runaway RSG is $\lesssim 0.02\%$. 

{\bf 2}: The RSG is the donor of the ULX and the offset of $66 \pm 6$ \vel{} is the systemic velocity of the binary: the BH (or neutron star) received a large natal kick (NK) and dragged the RSG (or its progenitor) with it (\citealt{repetto12}). 
Some Galactic BH binaries show high peculiar velocities (cf. \citealt{miller-jones14}). However, these are all BHs with low-mass companions. Instead, Cyg X-1, a binary containing a stellar mass BH with an $\sim 30$ \msun{} donor star (which is of the same magnitude as the RSG mass), has a peculiar velocity of only $\sim 20$ km/s.
The difficulty with this scenario is that the probability of kicking a BH/NS plus 10 \msun{} donor in a wide orbit to a high systemic velocity, while keeping the binary intact, is very small.

In order to quantify this we simulate a population of binaries containing the progenitor of the BH with a 10 \msun{} star as companion, calculating the effect of the BH formation event on the orbital and kinematical properties of the binary. The distribution of initial binary separations is taken to be flat between a minimum value $a_\textrm{min}$, corresponding to the orbital separation at which either one of the two components fills its Roche lobe, and a maximum value of $1000~a_\textrm{min}$. We model the BH formation event assuming a mass equal to half the mass of the BH is ejected instantaneously from the binary. In the most standard formation scenario, BHs are thought to be produced via fallback on to the nascent neutron stars. In such a scenario, BHs are then expected to receive a NK, as neutron stars do.
The NK will be however reduced by the ratio $M_\text{NS}/M_\text{BH}$, simply by conservation of linear momentum: $V_\text{NK, BH}=V_{\text{NK, NS}}(M_\text{NS}/M_\text{BH})$, where $V_\text{NK, NS}$ is drawn from a Maxwellian distribution as in \citet{hobbs05}. 
The NK has a random orientation with respect to the binary orbital plane.
Through the laws of conservation of energy and orbital angular momentum, we calculate the orbital parameters $(a, e)$ of the binary right after the BH is formed. We select among those binaries which stay bound in the process, those ones which satisfy two constraints.

The first constraint is that the systemic velocity acquired by the binary in the BH formation event, i.e. the velocity of the new centre of mass with respect to the old one, has to be larger than $60$ \vel.
The second constraint comes from the spectral and photometric properties of the companion star. The orbital separation after the BH is formed has to allow for mass transfer to happen on the nuclear time-scale of the companion when it fills its Roche lobe as an RSG. This is done making sure that the Roche lobe radius of the companion in the circularized orbit is in the interval $600-1600~R_{\odot}$.

\begin{figure}
\includegraphics[width=0.5\textwidth]{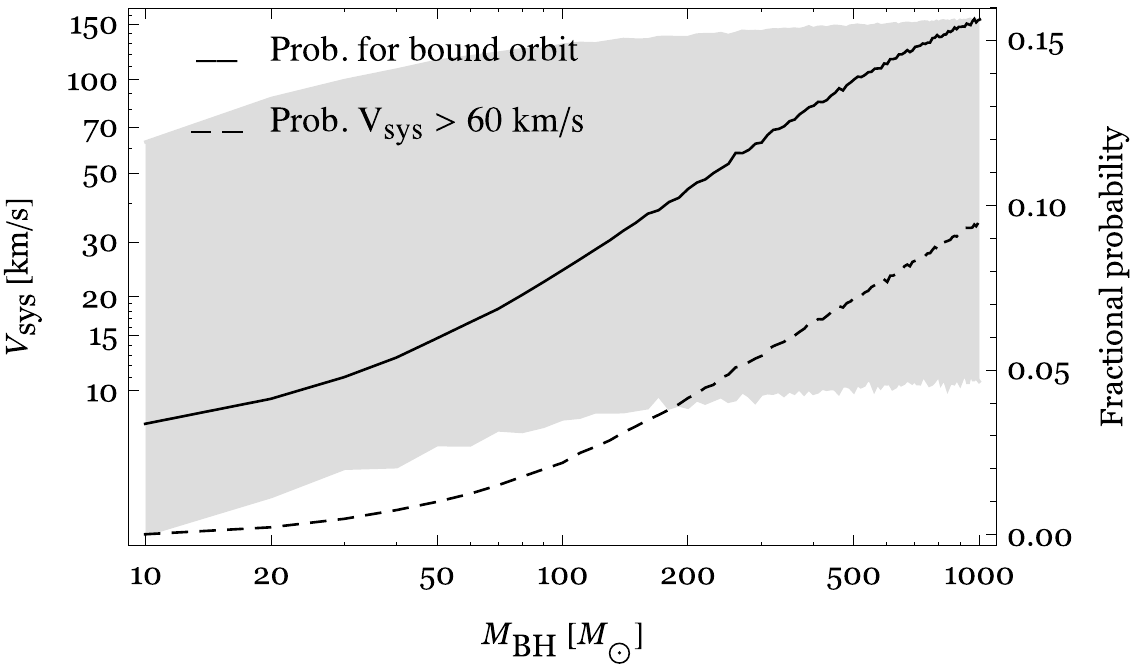}
\caption{Probability fraction for the binary to stay bound in the BH formation event (solid line),
and probability fraction that such a surviving binary acquires a systemic velocity greater then $60$ \vel{} (dashed line),
both as a function of the BH mass.
The shaded area corresponds to the range of possible systemic velocities acquired by the binary as a function
of the BH mass.}\label{fig:test}
\end{figure}

We show the results of this simulation in Figure \ref{fig:test}. The solid line shows the probability for the binary to stay bound in the BH formation event. The dashed line shows the probability for such a surviving binary to acquire a systemic velocity larger than $60$ \vel. Both probabilities are shown as a function of the BH mass. We also show in the plot the range of possible systemic velocities acquired by the surviving binary; the minimum and maximum value for the systemic velocity are calculated at the $99.7\%$ probability.
These simulations were done conservatively assuming an RSG of 10 \msun; if the donor star is more massive, the probabilities will be lower. 

Projection effects lower the probability even further. The chance that v$_{\textrm{sys}}$ is pointed within a degree from our line of sight is of the order of $10^{-5}$, so the space velocity of the RSG with respect to its environment is likely larger than what we measure in the radial direction.
Bearing in mind that BH formation is still very uncertain, we also tested a more conservative formation scenario for the BH, in which the BH does not receive any NK at birth, and the peculiar velocity at birth is a consequence of the mass ejection only (Blaauw kick, \citealt{blaauw61}).
In such a case, none of the surviving binaries acquire a velocity greater than $60$ km/s.

These simulations show that if the velocity difference is due to the systemic velocity only, a BH of $\gtrsim 50$ \msun{} is necessary to explain the measured offset.

{\bf 3}: The RSG is the donor of the ULX and (part of) the offset of 66 \vel{} is due to the binary motion of the RSG around the BH. 
It is difficult to explain the offset as the systemic velocity of the binary, but the velocity amplitude of an RSG orbiting a BH can be hundreds \vel. This requires a BH mass of $\gtrsim 100$ \msun{} for a system inclination $\leq 60^{\circ}$ (see Figure \ref{veloamp}).

Such a high BH mass seems at odds with what is expected based on analysis of the X-ray spectra of this ULX. However, \citet{pintore14} suggest this ULX may be viewed at high inclination, which would mean a lower mass BH is sufficient to explain the RV offset (Figure \ref{veloamp} shows the RV amplitude assuming an inclination of 60$^\circ$). If beaming plays a role in creating the high X-ray luminosities observed in ULXs, a high inclination might explain the relatively low X-ray luminosity of this source. The combination of these two effects might be enough to reconcile the two mass estimates.

With the current data we cannot distinguish among these scenarios for \j0047. New, high quality (X-shooter) spectra will allow us to measure radial velocity shifts if they are present. This will allow us to decide whether this ULX is a good target for dynamical measurements of its BH mass.

\section*{Acknowledgements}
We thank the anonymous referee for their comments that helped improve the paper. We want to thank Ben Davies for sharing his X-shooter RSG spectra and Tom Marsh for developing {\sc molly}. This paper uses data products produced by the OIR Telescope Data Center, supported by the Smithsonian Astrophysical Observatory. Based on observations made with ESO Telescopes at the La Silla Paranal Observatory under programme ID 093.D-0256. TPR's contributions were funded as part of the STFC consolidated grant award ST/L00075X/1.

\bibliographystyle{mn_new}
\bibliography{bibliography}

\end{document}